\documentclass[sigconf]{acmart}
\AtBeginDocument{%
  }

\setcopyright{acmlicensed}
\copyrightyear{2025}
\acmYear{2025}
\acmDOI{}
\acmConference[SIGIR]{ACM's Special Interest Group on Information Retrieval}{July 13-17, 2025}{Padova, Italy}
\acmISBN{978-1-4503-XXXX-X/2025/07}

\usepackage{graphicx}

\begin{document}

\title{ARAG: Agentic Retrieval Augmented Generation for \\ Personalized Recommendation}


\author{Reza Yousefi Maragheh}
\authornote{These authors contributed equally to this work.}
\email{Reza.YousefiMaragheh@walmart.com}
\affiliation{%
  \institution{Walmart Global Tech}
  \city{Sunnyvale}
  \state{California}
  \country{USA}
}

\author{Pratheek Vadla}
\authornotemark[1]
\email{Pratheek.Vadla@walmart.com}
\affiliation{%
  \institution{Walmart Global Tech}
  \city{Bellevue}
  \state{Washington}
  \country{USA}
}

\author{Priyank Gupta}
\authornotemark[1]
\email{Priyank.Gupta@walmart.com}
\affiliation{%
  \institution{Walmart Global Tech}
  \city{Bellevue}
  \state{Washington}
  \country{USA}
}

\author{Kai Zhao}
\authornotemark[1]
\email{Kai.Zhao@walmart.com}
\affiliation{%
  \institution{Walmart Global Tech}
  \city{Sunnyvale}
  \state{California}
  \country{USA}
}

\author{Aysenur Inan}
\authornotemark[1]
\email{Aysenur.Inan@walmart.com}
\affiliation{%
  \institution{Walmart Global Tech}
  \city{Sunnyvale}
  \state{California}
  \country{USA}
}

\author{Kehui Yao}
\authornotemark[1]
\email{Kehui.Yao@walmart.com}
\affiliation{%
  \institution{Walmart Global Tech}
  \city{Bellevue}
  \state{Washington}
  \country{USA}
}

\author{Jianpeng Xu}
\email{Jianpeng.Xu@walmart.com}
\affiliation{%
  \institution{Walmart Global Tech}
  \city{Sunnyvale}
  \state{California}
  \country{USA}
}

\author{Praveen Kanumala}
\email{Praveen.Kanumala@walmart.com}
\affiliation{%
  \institution{Walmart Global Tech}
  \city{Sunnyvale}
  \state{California}
  \country{USA}
}

\author{Jason Cho}
\email{Jason.Cho@walmart.com}
\affiliation{%
  \institution{Walmart Global Tech}
  \city{Sunnyvale}
  \state{California}
  \country{USA}
}

\author{Sushant Kumar}
\email{Sushant.Kumar@walmart.com}
\affiliation{%
  \institution{Walmart Global Tech}
  \city{Sunnyvale}
  \state{California}
  \country{USA}
}


\renewcommand{\shortauthors}{Maragheh et al.}

\begin{abstract}
Retrieval-Augmented Generation (RAG) has shown promise in enhancing recommendation systems by incorporating external context into large language model prompts. However, existing RAG-based approaches often rely on static retrieval heuristics and fail to capture nuanced user preferences in dynamic recommendation scenarios. In this work, we introduce ARAG, an Agentic Retrieval-Augmented Generation framework for Personalized Recommendation, which integrates a multi-agent collaboration mechanism into the RAG pipeline. To better understand the long-term and session behavior of the user, ARAG leverages four specialized LLM-based agents: a User Understanding Agent that summarizes user preferences from long-term and session contexts, a Natural Language Inference (NLI) Agent that evaluates semantic alignment between candidate items retrieved by RAG and inferred intent, a context summary agent that summarizes the findings of NLI agent, and an Item Ranker Agent that generates a ranked list of recommendations based on contextual fit. We evaluate ARAG accross three datasets. Experimental results demonstrate that ARAG significantly outperforms standard RAG and recency-based baselines, achieving up to 42.1\% improvement in NDCG@5 and 35.5\% in Hit@5. We also, conduct an ablation study to analyse the effect by different components of ARAG. Our findings highlight the effectiveness of integrating agentic reasoning into retrieval-augmented recommendation and provide new directions for LLM-based personalization.
\end{abstract}


\maketitle


\section{Introduction}
\begin{figure*}[h]
    \centering
    \includegraphics[width=0.97\linewidth]{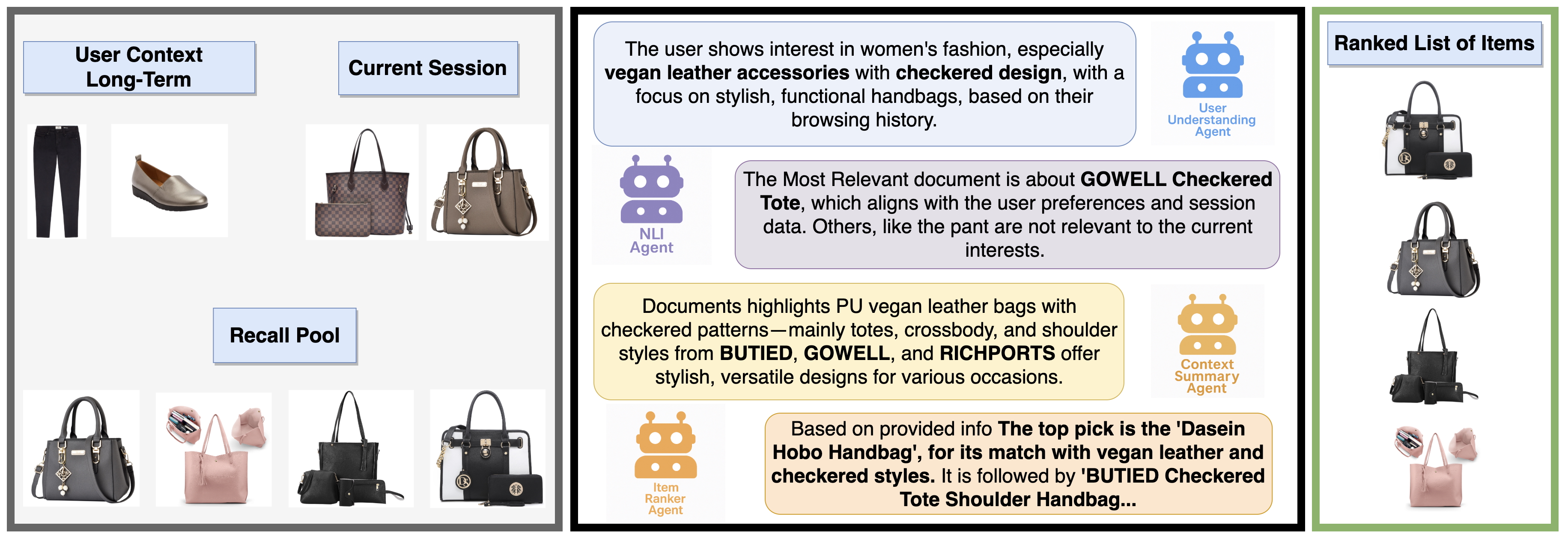}
    \caption{ARAG framework: the User Understanding Agent summarizes long-term and session-level preferences; the NLI Agent scores each candidate for semantic alignment; the Context Summary Agent condenses NLI-filtered evidence into a focused context; and the Item Ranker Agent fuses these signals to produce the final personalized ranking.}
    \label{figure:agentic_rag}
\end{figure*}

Adapting Retrieval-Augmented Generation (RAG) systems to recommendation scenarios offers promising opportunities to enhance the accuracy and personalization of suggestions~\cite{di2023retrieval, deldjoo2024review}. In this context, RAG can be leveraged to augment traditional recommendation algorithms with real-time, diverse information retrieval. By incorporating RAG, recommendation systems can go beyond relying solely on user preferences and item characteristics stored in a static database. Instead, they can dynamically fetch and consider additional relevant data such as recent trends, user reviews, expert opinions, or even real-time market data~\cite{su2024dragin}. This approach allows for more context-aware and up-to-date recommendations. For instance, a movie recommendation system using RAG could retrieve and analyze recent critic reviews, audience reactions, and current cultural trends to provide more timely and relevant suggestions. Moreover, RAG can help explain recommendations by retrieving and presenting supporting information, enhancing user trust and engagement~\cite{li2025g}. The adaptive nature of RAG also enables recommendation systems to handle long-tail items or new users more effectively by drawing upon a broader knowledge base, potentially addressing common challenges like cold start problems in collaborative filtering~\cite{wu2024coral}.

While RAG systems have shown promise in enhancing recommendation systems, there is significant room for improvement in their performance and effectiveness. The current limitations of RAG in recommendation contexts stem largely from their reliance on simplistic retrieval mechanisms, such as cosine similarity-based retrieval and embedding matching. These methods, while computationally efficient, often fall short in capturing the nuanced preferences and contexts that drive user behavior in recommendation scenarios~\cite{rossi2024relevance,mortaheb2025re}. The complexity of user preferences and the multifaceted nature of items being recommended demand more sophisticated approaches to information retrieval and matching.

A key area for advancement lies in developing RAG systems capable of better understanding and utilizing long-form user documents to infer user context~\cite{cui2024context, gunathilaka2025addressing, rakkappan2019context}. This involves moving beyond surface-level text matching to comprehend the implicit preferences, interests, and intentions embedded within user-generated content. Additionally, the ranking of the recall set of potential items requires more advanced algorithms that can weigh multiple factors simultaneously, including relevance, diversity, novelty, and contextual appropriateness~\cite{yu2024rankrag, ke2024bridging, maragheh2022prospect, ye2023seller, yousefi2020choice}. Future RAG systems for recommendations should incorporate more nuanced semantic understanding, and temporal dynamics to create a more holistic view of user preferences and item characteristics \cite{hong2023metagpt, chen2024magicore}. By addressing these challenges, RAG systems can evolve to provide more accurate, personalized, and contextually relevant recommendations, significantly enhancing the user experience across various recommendation platforms.

This paper presents ARAG: An Agentic Retrieval Augmented Generation framework for personalized recommendation. Building upon recent advancements in Agentic RAG across various domains, ARAG extends these principles to address the specific challenges of personalized recommendation systems. More specifically, under ARAG, and to better conduct the conversational ranking task, user long-term behavior is retrieved and passed through different agents to better adapt the rankings of documents.  

Through extensive experimentation, we assess the framework's efficacy in improving these key aspects of recommendation systems. The results demonstrate the potential of integrating agentic frameworks within the RAG paradigm for personalized recommendations, offering insights into how such an approach can enhance the accuracy and relevance of recommendations in various scenarios. 

\section{Methodology}

To better model user intent in personalized recommendation, we propose a multi-agent framework, ARAG, which introduces a set of specialized large language model (LLM) agents to refine context retrieval and item ranking. As shown in Figure~\ref{figure:agentic_rag}, the input to the system consists of two components: (1) a long-term context capturing the user’s historical interactions, and (2) the current session, reflecting recent user behaviors. These interaction histories are used to retrieve a set of semantically relevant candidate items through embedding-based similarity. However, instead of relying solely on static retrieval scores, ARAG applies a reasoning-oriented agentic workflow to assess the contextual alignment of each candidate item.

The workflow begins with the applying a regular RAG for retrieving the an initial set of larger recall set of items. Then, an NLI Agent evaluates whether each candidate item supports or aligns with this inferred user intent by analyzing its textual metadata (e.g., title, description, reviews). Then, a context summary agent summarizes the retrieved context by NLI agent. In parallel to this workflow, the User Understanding Agent, generates a natural language summary of user preferences based on the session context. This helps with identifying generic long-term interests of the user. Finally, the Item Ranker Agent integrates these signals to produce a final ranked list of items, prioritizing those most relevant to both the user’s current and historical preferences. This agentic collaboration enables ARAG to perform fine-grained relevance assessment and produce recommendations that are both context-aware and semantically grounded.

\subsection{Formal Problem Statement}
To formalize the methodology, the input to the system consists of two components: (i) A \textbf{long-term context}, $C_{\mathrm{lt}}$, capturing the user's historical interactions, and (ii) A \textbf{current session}, $C_{\mathrm{st}}$, reflecting recent user behaviors.

We let
\[
    \mathbf{u} = \bigl(C_{\mathrm{lt}}, \; C_{\mathrm{st}}\bigr)
\]
denote the combined user context. 

Let $\mathcal{I} = \{ i_1, \ldots, i_N\}$ be the set of all candidate items, each having associated textual metadata $T(i)$ (e.g., title, description, reviews). Our goal is to produce a final ranking, or a permutation $\pi$ over the $\mathcal{I}$:
\[
    \pi = f_{\mathrm{Rank}}(\mathbf{u}, \mathcal{I}),
\]
which orders items by their relevance to the user's context $\mathbf{u}$.
\subsection{ARAG framework}
In this subsection we formally introduce the components of ARAG. 

\subsubsection{Initial cosine similarity-based RAG} We use a RAG framework to obtain an initial subset $\mathcal{I}^0 \subseteq \mathcal{I}$ of candidate items. Assume there is an embedding function:
\[
    f_{\mathrm{Emb}}: \bigl(\mathcal{I} \cup \{\mathbf{u}\}\bigr) \rightarrow \mathbb{R}^d,
\]
which maps both items and user context into a shared $d$-dimensional embedding space. We measure similarity between two embeddings via $\mathrm{sim}(\cdot,\cdot)$, e.g., cosine similarity. 

The top-$k$ retrieved items are chosen by:
\[
    \mathcal{I}^0 
    = \mathrm{argtop}_{k} 
    \Bigl\{\,
    \mathrm{sim}\bigl(f_{\mathrm{Emb}}(i), f_{\mathrm{Emb}}(\mathbf{u})\bigr) 
    \;\Big|\; i \in \mathcal{I}
    \Bigr\}.
\]
This yields an initial \textit{recall set} of size $k$ that will be refined by subsequent agents.

\subsubsection{NLI Agent for Contextual Alignment} A \textbf{Natural Language Inference (NLI) Agent} evaluates each item $i \in \mathcal{I}^0$ to check how well its metadata $T(i)$ aligns with the user context. Let
\[
    s_{\mathrm{NLI}}(i, \mathbf{u}) 
    = \Phi\bigl(T(i), \mathbf{u}\bigr),
\]
denote the alignment score produced by the NLI Agent, where $\Phi$ is an LLM-based function. A high score indicates that $i$ strongly supports or matches the user’s interests.

\subsubsection{Context Summary Agent} A \textbf{Context Summary Agent} (CSA) then summarizes the textual metadata of only those candidate items that the NLI Agent has deemed sufficiently aligned with the user context. Formally, define
\[
    \mathcal{I}^{+} 
    = \{\, i \in \mathcal{I}^{0} \;\mid\; s_{\mathrm{NLI}}(i, \mathbf{u}) \ge \theta \},
\]
where \(s_{\mathrm{NLI}}(i, \mathbf{u})\) is the NLI alignment score and \(\theta\) is a threshold above which an item is considered accepted. The Context Summary Agent then produces a concise summary,
\[
    S_{\mathrm{ctx}} 
    = \Psi\Bigl\{ T(i) \;\Big|\; i \in \mathcal{I}^{+} \Bigr\},
\]
where \(\Psi(\cdot)\) is an LLM-driven summarization function operating on the textual metadata \(T(i)\) of each accepted item.

\subsubsection{User Understanding Agent} In parallel, the \textbf{User Understanding Agent} (UUA) synthesizes a high-level summary of the user’s preferences, based on the long-term context $C_{\mathrm{lt}}$ and the current session $C_{\mathrm{st}}$:
\[
    S_{\mathrm{user}} 
    = \Omega\bigl(\mathbf{u}\bigr),
\]
where $\Omega(\cdot)$ is an LLM-based reasoning function that generates a natural language description of the user’s generic interests and immediate goals.

\subsubsection{Item Ranker Agent}

Finally, the \textbf{Item Ranker Agent} (IRA) uses outputs $S_{\mathrm{user}}$ and $S_{\mathrm{ctx}}$ as context for ranking. Prompt for Ranker agent explicitly instructs the model to: (1) consider the user’s behavior in previous sessions, (2) consider the relevant part of the user history to the current ranking task, (3) examine the candidate items, and (4) rank the items in descending order of purchase likelihood. For example, given a user summary indicating interest in vegan leather products, checkered bags, and stylish accessories, the ranker may prioritize items like the BUTIED Checkered Tote Shoulder Handbag over the Dasein Hobo Handbag and Women's Large Tote based on alignment with both material and style preferences.

Formally, the model returns a permutation over the $r$ candidate items:
\[
\pi = f_{\text{rank}}(S_{\mathrm{user}}, S_{\mathrm{ctx}} , \mathcal{I}), \quad \pi = \{r_1, r_2, \dots, r_N\}
\]
where each $r_j \in \{1, 2, \dots, N\}$ denotes the index of item in rank $j$ in the final ranked list. 

As one can see under ARAG NLI, context summary, and user understanding agents collectively act as a memory moderation scheme for the final ranking task. These agents are utilized to make sure the user's long-term and short-term behavioral context is properly integrated into the final ranking task. 

\subsection{Agent Collaboration Protocol}
To better explain the implementation workflow, we also clarify the collaboration protocol for agents in ARAG. ARAG is implemented as a \emph{blackboard‐style} multi‑agent system \cite{nii1986blackboard} in which all agents read from and write to a shared, structured memory $\mathcal{B}$. Each agent contributes a \texttt{message} object \texttt{m} containing a JSON schema
\{\texttt{id}, \texttt{role}, \texttt{content}, \texttt{score}, \texttt{timestamp}\}, so that subsequent agents can reason not only over the raw user and item data, but also over the rationales produced by their peers.

\begin{table*}[h]
\centering
\begin{tabular}{l|cc|cc|cc}
\hline
\multicolumn{7}{c}{\textbf{Performance of Benchmark Versus ARAG}} \\
\hline
& \multicolumn{2}{c|}{ Clothing} & \multicolumn{2}{c|}{ Electronics} & \multicolumn{2}{c}{ Home} \\
& NDCG@5 & Hit@5 & NDCG@5 & Hit@5 & NDCG@5 & Hit@5 \\
\hline
Recency-based Ranking & 0.30915 & 0.3945 & 0.22482 & 0.3035 & 0.22443 & 0.2988 \\
Vanilla RAG & 0.29884 & 0.3792 & 0.23817 & 0.321 & 0.22901 & 0.3117 \\
\hline
Agentic RAG & \textbf{0.43937} & \textbf{0.5347} & \textbf{0.32853} & \textbf{0.4201} & \textbf{0.28863} & \textbf{0.3834} \\
\% Improvement & 42.12\% & 35.54\% & 37.94\% & 30.87\% & 25.60\% & 22.68\% \\
\hline
\hline
\multicolumn{7}{c}{\textbf{Ablation Study}} \\
\hline
& \multicolumn{2}{c|}{Clothing} & \multicolumn{2}{c|}{Electronics} & \multicolumn{2}{c}{Home} \\
& NDCG@5 & HIT@5 & NDCG@5 & HIT@5 & NDCG@5 & HIT@5 \\
\hline
Vanilla RAG & 0.29884 & 0.3792 & 0.23817 & 0.321 & 0.22901 & 0.3117 \\
ARAG w/o NLI \& CSA & 0.3024 & 0.3859 & 0.2724 & 0.3559 & 0.2494 & 0.3308 \\
ARAG w/o NLI & 0.3849 & 0.4714 & 0.296 & 0.3878 & 0.2732 & 0.3582 \\
ARAG & \textbf{0.43937} & \textbf{0.5347} & \textbf{0.32853} & \textbf{0.4201} & \textbf{0.28863} & \textbf{0.3834} \\
\hline
\end{tabular}
\caption{Performance comparison of ARAG with benchmark models and ablation study on components of ARAG on Amazon datasets. 
}
\label{tab:combined}
\end{table*}
\begin{enumerate}
    \item \textbf{Parallel inference.}  
    The \emph{User‑Understanding Agent} (UUA) and the \emph{NLI Agent} are executed \emph{concurrently}.  
    The UUA writes a preference summary $\mathbf{m}_{\textit{user}}$ to $\mathcal{B}$, while the NLI Agent
    writes a support/contradiction judgement vector 
    $\mathbf{m}_{\textit{nli}} = \bigl[s_{\mathrm{NLI}}(i,\mathbf{u})\bigr]_{i\in\mathcal{I}^0}$.

    \item \textbf{Cross‑agent attention.}  
    The \emph{Context‑Summary Agent} (CSA) attends to \emph{both}
    $\mathbf{m}_{\textit{user}}$ and $\mathbf{m}_{\textit{nli}}$:  
    it uses the user summary as a relevance prior and the NLI scores as salience weights
    when composing $S_{\mathrm{ctx}}$, which it then records as $\mathbf{m}_{\textit{ctx}}$.

    \item \textbf{Final Ranking.}  
    The \emph{Item‑Ranker Agent} (IRA) consumes \{\,$\mathbf{m}_{\textit{user}}$, $\mathbf{m}_{\textit{ctx}}$\,\}
    and generates a ranked list $\pi$ together with an
    \emph{explanation trace}.  
\end{enumerate}

The above steps provide a multi-agent, reasoning-oriented approach that refines an initial set of retrieved items into a contextually aligned recommendation list. By delegating specialized tasks (e.g., NLI, summarization, user understanding, and ranking) to different large language model agents, \textbf{ARAG} enables: (i) \textit{Context Awareness:} Both long-term and short-term user behaviors factor into the final ranking, (ii) \textit{Semantic Grounding:} NLI and summarization agents enhance interpretability and precision, and (iii) \textit{Personalization:} The final score reflects the user’s unique and evolving preferences, ensuring recommendations remain both relevant and adaptable.

\section{Experiments}
\subsection{Dataset}

Our experiments utilize the widely-adopted Amazon Review dataset (He \& McAuley, 2016), a large-scale collection of product reviews and metadata spanning multiple product categories on Amazon.com. This dataset contains millions of customer reviews, ratings, and product interactions across diverse categories including Electronics, Books, Clothing, and Home \& Kitchen, making it particularly suitable for evaluating cross-category recommendation performance. For our experiments, we selected a subset of user-item interactions from 10,000 randomly sampled users across these categories. Each review entry contains rich contextual information including timestamps, ratings, textual feedback, and product metadata, providing comprehensive user preference signals. This dataset presents realistic challenges for recommendation systems, including sparse interaction matrices, shifting user preferences over time, and diverse product taxonomies—making it an ideal testbed for evaluating the ARAG framework's ability to leverage complex user contexts.

\subsection{Benchmark Models}

The Recency model adopts a simple temporal heuristic, assuming that a user's most recent interactions best reflect current preferences. It appends these recent items directly to the LLM prompt without further filtering. This model operationalizes recency by directly appending the user's most recent historical interactions to the large language model's input prompt, without additional filtering or transformation mechanisms. By prioritizing chronologically recent user behavior over potentially more relevant but temporally distant interactions, this approach benefits from simplicity and computational efficiency.

The Vanilla RAG (Retrieval-Augmented Generation) approach implements a more sophisticated information retrieval mechanism that moves beyond simple temporal ordering. This benchmark leverages embedding-based retrieval to identify semantically relevant items from the user's interaction history, selecting items based on their embedding similarity rather than recency. After retrieving these relevant historical items, the model appends them to the LLM prompt to provide context for generating recommendations. For all the experiments, we used gpt-3.5-turbo (v0125), and set the temperature argument to $0$ to increase repeatability of the experiments.

\subsection{Results}
NDCG@5 and HIT@5 scores for ARAG as well as benchmark models is presented in the first part of Table \ref{tab:combined}. The results demonstrate that the Agentic RAG approach significantly outperforms both Recency-based Ranking and Vanilla RAG frameworks across all datasets and metrics. Examining the NDCG@5 scores, Agentic RAG achieves 0.439, 0.329, and 0.289 on Amazon Clothing, Electronics, and Home respectively, compared to the next best performing approach which ranges from 0.299 to 0.238. Similarly, Hit@5 metrics show consistent superiority with Agentic achieving 0.535, 0.420, and 0.383 across the three domains. These substantial improvements suggest that the agentic approach to retrieval provides a more effective mechanism for identifying and ranking relevant recommendations in conversational systems.

The improvement percentages quantify the magnitude of Agentic RAG's performance gains, showing the most dramatic enhancement in the Clothing domain (42.12\% for NDCG@5, 35.54\% for Hit@5), followed by Electronics (37.94\% and 30.87\%) and Home (25.60\% and 22.68\%). This pattern suggests that the effectiveness of Agentic RAG may vary by domain characteristics, with potentially greater benefits in categories where item attributes and user preferences are more diverse or complex. The consistency of improvement across all datasets validates that the agentic approach addresses fundamental limitations in both recency-based and standard retrieval methods for conversational recommendation tasks.

Interestingly, the comparative performance between Recency-based Ranking and Vanilla RAG varies by domain. In the Clothing category, Recency-based approaches outperform Vanilla RAG (0.309 vs. 0.299 for NDCG@5), while the opposite holds for Electronics and Home categories. This observation points to domain-specific dynamics in user recommendation patterns, where temporal recency may be more valuable in fashion-oriented categories compared to electronics or home goods. Despite these variations, the consistent superiority of Agentic RAG across all domains indicates that the intelligent, adaptive retrieval strategies employed by the agentic framework offer substantial advantages over both temporal and standard retrieval approaches in conversational recommendation scenarios.

\subsection{Ablation Study}

The ablation study (second part of Table~\ref{tab:combined}) highlights the incremental benefits of each component in our recommendation pipeline. Starting from the Vanilla RAG baseline, which yields modest NDCG@5 scores of 0.299, 0.238, and 0.229 for Clothing, Electronics, and Home, adding the User Summary Agent leads to consistent gains across all domains—most notably in Electronics (14.4\% improvement in NDCG@5) and Home (8.9\%). These results confirm the importance of user preference summarization for enhancing context relevance beyond static embedding-based retrieval.

Introducing the Context Summary Agent further boosts performance, especially in the Clothing domain (28.8\% NDCG@5 improvement), suggesting that item-level contextual understanding is critical in categories where compatibility and style matter. The complete Agentic RAG system, incorporating all components, achieves the best results, with up to 14\% additional gains in NDCG@5 for Clothing. This confirms that semantic reasoning via natural language inference effectively bridges the gap between user intent and candidate item representation. Together, the agents provide complementary value, enabling state-of-the-art performance in conversational recommendation.

\section{Conclusion}

ARAG reframes retrieval‑augmented recommendation as a coordinated reasoning task among four specialized LLM agents. By separating the concerns of user understanding, semantic alignment, context synthesis, and ranking, the framework converts a coarse embedding‑based recall set into a finely filtered, semantically grounded candidate list that directly reflects both long‑term preferences and session intent. Extensive experiments on three benchmarks show that this agentic decomposition yields substantial accuracy gains while simultaneously providing transparent rationales that enhance interpretability and user trust. These results demonstrate that agent‑oriented orchestration inside the RAG loop is an effective, practical route to highly personalized, context‑aware recommendation.

\bibliographystyle{ACM-Reference-Format}
\bibliography{sample-base}


\begin{thebibliography}{18}


\ifx \showCODEN    \undefined \def \showCODEN     #1{\unskip}     \fi
\ifx \showISBNx    \undefined \def \showISBNx     #1{\unskip}     \fi
\ifx \showISBNxiii \undefined \def \showISBNxiii  #1{\unskip}     \fi
\ifx \showISSN     \undefined \def \showISSN      #1{\unskip}     \fi
\ifx \showLCCN     \undefined \def \showLCCN      #1{\unskip}     \fi
\ifx \shownote     \undefined \def \shownote      #1{#1}          \fi
\ifx \showarticletitle \undefined \def \showarticletitle #1{#1}   \fi
\ifx \showURL      \undefined \def \showURL       {\relax}        \fi
\providecommand\bibfield[2]{#2}
\providecommand\bibinfo[2]{#2}
\providecommand\natexlab[1]{#1}
\providecommand\showeprint[2][]{arXiv:#2}

\bibitem[Chen et~al\mbox{.}(2024)]%
        {chen2024magicore}
\bibfield{author}{\bibinfo{person}{Justin Chih-Yao Chen}, \bibinfo{person}{Archiki Prasad}, \bibinfo{person}{Swarnadeep Saha}, \bibinfo{person}{Elias Stengel-Eskin}, {and} \bibinfo{person}{Mohit Bansal}.} \bibinfo{year}{2024}\natexlab{}.
\newblock \showarticletitle{MAgICoRe: Multi-Agent, Iterative, Coarse-to-Fine Refinement for Reasoning}.
\newblock \bibinfo{journal}{\emph{arXiv preprint arXiv:2409.12147}} (\bibinfo{year}{2024}).
\newblock


\bibitem[Cui et~al\mbox{.}(2024)]%
        {cui2024context}
\bibfield{author}{\bibinfo{person}{Ziqiang Cui}, \bibinfo{person}{Haolun Wu}, \bibinfo{person}{Bowei He}, \bibinfo{person}{Ji Cheng}, {and} \bibinfo{person}{Chen Ma}.} \bibinfo{year}{2024}\natexlab{}.
\newblock \showarticletitle{Context Matters: Enhancing Sequential Recommendation with Context-aware Diffusion-based Contrastive Learning}. In \bibinfo{booktitle}{\emph{Proceedings of the 33rd ACM International Conference on Information and Knowledge Management}}. \bibinfo{pages}{404--414}.
\newblock


\bibitem[Deldjoo et~al\mbox{.}(2024)]%
        {deldjoo2024review}
\bibfield{author}{\bibinfo{person}{Yashar Deldjoo}, \bibinfo{person}{Zhankui He}, \bibinfo{person}{Julian McAuley}, \bibinfo{person}{Anton Korikov}, \bibinfo{person}{Scott Sanner}, \bibinfo{person}{Arnau Ramisa}, \bibinfo{person}{Ren{\'e} Vidal}, \bibinfo{person}{Maheswaran Sathiamoorthy}, \bibinfo{person}{Atoosa Kasirzadeh}, {and} \bibinfo{person}{Silvia Milano}.} \bibinfo{year}{2024}\natexlab{}.
\newblock \showarticletitle{A review of modern recommender systems using generative models (gen-recsys)}. In \bibinfo{booktitle}{\emph{Proceedings of the 30th ACM SIGKDD Conference on Knowledge Discovery and Data Mining}}. \bibinfo{pages}{6448--6458}.
\newblock


\bibitem[Di~Palma(2023)]%
        {di2023retrieval}
\bibfield{author}{\bibinfo{person}{Dario Di~Palma}.} \bibinfo{year}{2023}\natexlab{}.
\newblock \showarticletitle{Retrieval-augmented recommender system: Enhancing recommender systems with large language models}. In \bibinfo{booktitle}{\emph{Proceedings of the 17th ACM Conference on Recommender Systems}}. \bibinfo{pages}{1369--1373}.
\newblock


\bibitem[Gunathilaka et~al\mbox{.}(2025)]%
        {gunathilaka2025addressing}
\bibfield{author}{\bibinfo{person}{Thennakoon Mudiyanselage Anupama~Udayangani Gunathilaka}, \bibinfo{person}{Prabhashrini~Dhanushika Manage}, \bibinfo{person}{Jinglan Zhang}, \bibinfo{person}{Yuefeng Li}, {and} \bibinfo{person}{Wayne Kelly}.} \bibinfo{year}{2025}\natexlab{}.
\newblock \showarticletitle{Addressing sparse data challenges in recommendation systems: A Systematic review of rating estimation using sparse rating data and profile enrichment techniques}.
\newblock \bibinfo{journal}{\emph{Intelligent Systems with Applications}} (\bibinfo{year}{2025}), \bibinfo{pages}{200474}.
\newblock


\bibitem[Hong et~al\mbox{.}(2023)]%
        {hong2023metagpt}
\bibfield{author}{\bibinfo{person}{Sirui Hong}, \bibinfo{person}{Xiawu Zheng}, \bibinfo{person}{Jonathan Chen}, \bibinfo{person}{Yuheng Cheng}, \bibinfo{person}{Jinlin Wang}, \bibinfo{person}{Ceyao Zhang}, \bibinfo{person}{Zili Wang}, \bibinfo{person}{Steven Ka~Shing Yau}, \bibinfo{person}{Zijuan Lin}, \bibinfo{person}{Liyang Zhou}, {et~al\mbox{.}}} \bibinfo{year}{2023}\natexlab{}.
\newblock \showarticletitle{Metagpt: Meta programming for multi-agent collaborative framework}.
\newblock \bibinfo{journal}{\emph{arXiv preprint arXiv:2308.00352}} \bibinfo{volume}{3}, \bibinfo{number}{4} (\bibinfo{year}{2023}), \bibinfo{pages}{6}.
\newblock


\bibitem[Ke et~al\mbox{.}(2024)]%
        {ke2024bridging}
\bibfield{author}{\bibinfo{person}{Zixuan Ke}, \bibinfo{person}{Weize Kong}, \bibinfo{person}{Cheng Li}, \bibinfo{person}{Mingyang Zhang}, \bibinfo{person}{Qiaozhu Mei}, {and} \bibinfo{person}{Michael Bendersky}.} \bibinfo{year}{2024}\natexlab{}.
\newblock \showarticletitle{Bridging the preference gap between retrievers and llms}.
\newblock \bibinfo{journal}{\emph{arXiv preprint arXiv:2401.06954}} (\bibinfo{year}{2024}).
\newblock


\bibitem[Li et~al\mbox{.}(2025)]%
        {li2025g}
\bibfield{author}{\bibinfo{person}{Yuhan Li}, \bibinfo{person}{Xinni Zhang}, \bibinfo{person}{Linhao Luo}, \bibinfo{person}{Heng Chang}, \bibinfo{person}{Yuxiang Ren}, \bibinfo{person}{Irwin King}, {and} \bibinfo{person}{Jia Li}.} \bibinfo{year}{2025}\natexlab{}.
\newblock \showarticletitle{G-Refer: Graph Retrieval-Augmented Large Language Model for Explainable Recommendation}.
\newblock \bibinfo{journal}{\emph{arXiv preprint arXiv:2502.12586}} (\bibinfo{year}{2025}).
\newblock


\bibitem[Maragheh et~al\mbox{.}(2022)]%
        {maragheh2022prospect}
\bibfield{author}{\bibinfo{person}{Reza~Yousefi Maragheh}, \bibinfo{person}{Ramin Giahi}, \bibinfo{person}{Jianpeng Xu}, \bibinfo{person}{Lalitesh Morishetti}, \bibinfo{person}{Shanu Vashishtha}, \bibinfo{person}{Kaushiki Nag}, \bibinfo{person}{Jason Cho}, \bibinfo{person}{Evren Korpeoglu}, \bibinfo{person}{Sushant Kumar}, {and} \bibinfo{person}{Kannan Achan}.} \bibinfo{year}{2022}\natexlab{}.
\newblock \showarticletitle{Prospect-net: Top-k retrieval problem using prospect theory}. In \bibinfo{booktitle}{\emph{2022 IEEE International Conference on Big Data (Big Data)}}. IEEE, \bibinfo{pages}{3945--3951}.
\newblock


\bibitem[Mortaheb et~al\mbox{.}(2025)]%
        {mortaheb2025re}
\bibfield{author}{\bibinfo{person}{Matin Mortaheb}, \bibinfo{person}{Mohammad A~Amir Khojastepour}, \bibinfo{person}{Srimat~T Chakradhar}, {and} \bibinfo{person}{Sennur Ulukus}.} \bibinfo{year}{2025}\natexlab{}.
\newblock \showarticletitle{Re-ranking the Context for Multimodal Retrieval Augmented Generation}.
\newblock \bibinfo{journal}{\emph{arXiv preprint arXiv:2501.04695}} (\bibinfo{year}{2025}).
\newblock


\bibitem[Nii(1986)]%
        {nii1986blackboard}
\bibfield{author}{\bibinfo{person}{H~Penny Nii}.} \bibinfo{year}{1986}\natexlab{}.
\newblock \showarticletitle{The blackboard model of problem solving and the evolution of blackboard architectures}.
\newblock \bibinfo{journal}{\emph{AI magazine}} \bibinfo{volume}{7}, \bibinfo{number}{2} (\bibinfo{year}{1986}), \bibinfo{pages}{38--38}.
\newblock


\bibitem[Rakkappan and Rajan(2019)]%
        {rakkappan2019context}
\bibfield{author}{\bibinfo{person}{Lakshmanan Rakkappan} {and} \bibinfo{person}{Vaibhav Rajan}.} \bibinfo{year}{2019}\natexlab{}.
\newblock \showarticletitle{Context-aware sequential recommendations withstacked recurrent neural networks}. In \bibinfo{booktitle}{\emph{The world wide web conference}}. \bibinfo{pages}{3172--3178}.
\newblock


\bibitem[Rossi et~al\mbox{.}(2024)]%
        {rossi2024relevance}
\bibfield{author}{\bibinfo{person}{Nicholas Rossi}, \bibinfo{person}{Juexin Lin}, \bibinfo{person}{Feng Liu}, \bibinfo{person}{Zhen Yang}, \bibinfo{person}{Tony Lee}, \bibinfo{person}{Alessandro Magnani}, {and} \bibinfo{person}{Ciya Liao}.} \bibinfo{year}{2024}\natexlab{}.
\newblock \showarticletitle{Relevance filtering for embedding-based retrieval}. In \bibinfo{booktitle}{\emph{Proceedings of the 33rd ACM International Conference on Information and Knowledge Management}}. \bibinfo{pages}{4828--4835}.
\newblock


\bibitem[Su et~al\mbox{.}(2024)]%
        {su2024dragin}
\bibfield{author}{\bibinfo{person}{Weihang Su}, \bibinfo{person}{Yichen Tang}, \bibinfo{person}{Qingyao Ai}, \bibinfo{person}{Zhijing Wu}, {and} \bibinfo{person}{Yiqun Liu}.} \bibinfo{year}{2024}\natexlab{}.
\newblock \showarticletitle{DRAGIN: Dynamic Retrieval Augmented Generation based on the Information Needs of Large Language Models}.
\newblock \bibinfo{journal}{\emph{arXiv preprint arXiv:2403.10081}} (\bibinfo{year}{2024}).
\newblock


\bibitem[Wu et~al\mbox{.}(2024)]%
        {wu2024coral}
\bibfield{author}{\bibinfo{person}{Junda Wu}, \bibinfo{person}{Cheng-Chun Chang}, \bibinfo{person}{Tong Yu}, \bibinfo{person}{Zhankui He}, \bibinfo{person}{Jianing Wang}, \bibinfo{person}{Yupeng Hou}, {and} \bibinfo{person}{Julian McAuley}.} \bibinfo{year}{2024}\natexlab{}.
\newblock \showarticletitle{Coral: Collaborative retrieval-augmented large language models improve long-tail recommendation}. In \bibinfo{booktitle}{\emph{Proceedings of the 30th ACM SIGKDD Conference on Knowledge Discovery and Data Mining}}. \bibinfo{pages}{3391--3401}.
\newblock


\bibitem[Ye et~al\mbox{.}(2023)]%
        {ye2023seller}
\bibfield{author}{\bibinfo{person}{Zikun Ye}, \bibinfo{person}{Reza~Yousefi Maragheh}, \bibinfo{person}{Lalitesh Morishetti}, \bibinfo{person}{Shanu Vashishtha}, \bibinfo{person}{Jason Cho}, \bibinfo{person}{Kaushiki Nag}, \bibinfo{person}{Sushant Kumar}, {and} \bibinfo{person}{Kannan Achan}.} \bibinfo{year}{2023}\natexlab{}.
\newblock \showarticletitle{Seller-side Outcome Fairness in Online Marketplaces}.
\newblock \bibinfo{journal}{\emph{arXiv preprint arXiv:2312.03253}} (\bibinfo{year}{2023}).
\newblock


\bibitem[Yousefi~Maragheh et~al\mbox{.}(2020)]%
        {yousefi2020choice}
\bibfield{author}{\bibinfo{person}{Reza Yousefi~Maragheh}, \bibinfo{person}{Xin Chen}, \bibinfo{person}{James Davis}, \bibinfo{person}{Jason Cho}, \bibinfo{person}{Sushant Kumar}, {and} \bibinfo{person}{Kannan Achan}.} \bibinfo{year}{2020}\natexlab{}.
\newblock \showarticletitle{Choice modeling and assortment optimization in the presence of context effects}.
\newblock \bibinfo{journal}{\emph{Available at SSRN 3747354}} (\bibinfo{year}{2020}).
\newblock


\bibitem[Yu et~al\mbox{.}(2024)]%
        {yu2024rankrag}
\bibfield{author}{\bibinfo{person}{Yue Yu}, \bibinfo{person}{Wei Ping}, \bibinfo{person}{Zihan Liu}, \bibinfo{person}{Boxin Wang}, \bibinfo{person}{Jiaxuan You}, \bibinfo{person}{Chao Zhang}, \bibinfo{person}{Mohammad Shoeybi}, {and} \bibinfo{person}{Bryan Catanzaro}.} \bibinfo{year}{2024}\natexlab{}.
\newblock \showarticletitle{Rankrag: Unifying context ranking with retrieval-augmented generation in llms}.
\newblock \bibinfo{journal}{\emph{Advances in Neural Information Processing Systems}}  \bibinfo{volume}{37} (\bibinfo{year}{2024}), \bibinfo{pages}{121156--121184}.
\newblock


\end{thebibliography}


\end{document}